\documentstyle[12pt,a4]{article}

\input {epsf}

\def\beq{\begin{equation}}
\def\eeq{\end{equation}}
\def\etal{et al. }
\def\ie{i.e.}
\def\surr{{\mbox{\small surr}}}
\def\Per{\mbox{Per}}
\def\ACF{{\bf R}}

\def\x{{\bf x}}

\begin{document}

\title      {   On Surrogate Data Testing for Linearity \\
                based on the Periodogram
                }
\author {       J. Timmer \\
                Fakult\"at f\"ur Physik \\
                Herman -- Herder -- Str. 3 \\
                79104 Freiburg \\
                Germany
                         }
\date{}
\maketitle
\begin {abstract}
The method of surrogate data is a tool to test whether data were
generated by some class of model. Tests based on the periodogram
have been proposed to decide if linear systems driven by Gaussian
noise could have generated a sample time series. We show that this
procedure based on the periodogram, in general, misspecifies the test
statistic. Based on the theory of linear systems we suggest an
alternative procedure to obtain the correct distribution of the test
statistic and discuss problems of this approach.

\end {abstract}
%

\clearpage
\section {Introduction}

The method of surrogate data was proposed by Theiler \etal
\cite{theiler1} as a general procedure to test whether data are
consistent with some class of models. It has been used to test a
proposed nonlinear dynamics in geophysics \cite {witt},
 certain properties of neural
spike dynamics \cite {longtin}, but the most common application is
testing for linear models
\cite{grassberger,kurths,fraser,lopes,schiff,theiler2,palus}.
In order to test the hypothesis that the data are consistent with
being generated by a linear system, the FT algorithm is applied.
Based on an example and  the theory of linear stochastic systems we
 will show that
this algorithm does not produce the correct distribution of time
series and therefore might not generally yield the correct distribution of
 the test statistic.

The surrogate data method is similar to a Monte Carlo implementation
of a standard hypothesis test in which one specifies a procedure for
generating an ensemble of new time series with a distribution that
matches a model to be tested.  One generates several sample time
series, the surrogate data.  Next, a selected feature function
$f$ is evaluated for the original time series $\x$ and for each of the
surrogates. If the distribution of the feature values is well approximated by
a Gaussian it can be characterized by its  mean $\mu_{\surr}$ and
 standard deviation
$\sigma_{\surr}$.  If $f(\x)$, the feature value calculated for the
measured time series, is more than a few standard deviations away from
$\mu_\surr$, the null hypothesis that the original data were generated
by the tested model is rejected, and the significance is reported as the
distance $z=\frac{|\mu_\surr - f(\x)|}{\sigma_\surr}$. If the distribution
of the feature is not well approximated by a Gaussian distribution,
nonparametric rank statistics  should be used.

The surrogate data method differs from a simple Monte Carlo
implementation of a hypothesis test in that it tests not against a
single model, but a class of models, i.e. linear systems driven by Gaussian
noise.
The idea is to select a single  model from the class on the basis of
the measured data $\x$, and then do a Monte Carlo hypothesis test for the
selected model and the original data.

 The statistical properties of a time
series generated by a linear process are specified by the
autocovariance function (ACF) or equivalently by its Fourier
transform, the power spectrum $ S (\omega) $.  Theiler
\etal use the periodogram $ \Per (\omega) $, \ie, the squared modulus
of the Fourier transform of the data, to specify the linear model to be
 tested\footnote{An invertible nonlinearity
in the observation function should be rectified by a change of coordinates
that forces the distribution of the data to be Gaussian \cite {rapp}.}.
They generate surrogates (FT algorithm) by drawing complex numbers with
amplitude $\sqrt{\Per (\omega)} $ and random phases and Fourier
transforming back into the time domain.
There are two problems in this procedure: First, the periodogram
$\Per(\omega)$ is not a consistent estimator of the power spectrum
$S(\omega)$, and second, the distribution of surrogates generated by
randomizing the phases but not the amplitudes does not correspond to
any linear stochastic system at all. The consequences of these two points on
the estimated distribution of the feature,
in case of Gaussianity $\mu_{\surr}$ and $\sigma_{\surr} $,
are the primary focus of this paper.

In the next section we briefly summarize the main
ideas of surrogate data testing for linearity based on the periodogram
and demonstrate by a simple example that this procedure might  misspecificy
the distribution of the test statistic.
To explain this result we refer to
some results of the theory of linear systems in section
\ref{review} and discuss the consequences for surrogate data
testing for linearity in section \ref{consequ}.

\section{Periodogram Based Surrogates} \label{surro}

The key to a hypothesis test is the distribution of a  test
statistic $f$ when the null hypothesis $H_0$ is true,
i.e. $p(f|H_0)$.  Given a measured sample ${\bf{x}}$, one calculates
$ f(\bf{x}) $ and decides if it is reasonable to suppose
that $f(\bf{x}) $ was drawn from $p(f|H_0)$, in general by an a priori
choosen significance level.

The surrogate data method was proposed by Theiler \etal primarily to
test the null hypothesis that the data come from a linear Gaussian process.
 Once again, the key to designing a hypothesis test is the
identification of the distribution of a test statistic $p(f|H_0)$.
  For a class as large and
complex as all linear processes, it seems impossible.  The idea of
Theiler \etal is to use the observation itself to limit the
size of the whole class to a single member.

Surrogate data testing needs two ingredients: A procedure
to generate data from the class of models to be tested and a feature
or statistic to perform the testing. If a single model is given, the
first step is simple.  One generates a large number of time series and
approximates the distribution $p(f|H_0)$.  For instance, see the
paper by Witt
\etal \cite {witt} in which a nonlinear difference equation proposed to
describe the dynamics of the reversals of the earth's magnetic field
was rejected using a feature that describes the transition
probabilities.

Theiler \etal proposed their FT algorithm to determine if a
given time series could have been produced by a linear system driven
by Gaussian noise.
The basic idea is that the statistical properties of a linear process
are captured completely by its spectral properties.
 Therefore, they first
calculate the periodogram $\Per(\omega) $, \ie the squared modulus of
the Fourier transform, for the positive frequencies.  In order to
generate the surrogate data they draw complex numbers with amplitude $
\sqrt{\Per (\omega)} $ and random phases, symmetrize the negative
frequencies with $ x(-\omega)=x(\omega)^{\ast} $ to obtain a real time
series by Fourier transforming back into the time domain.

We now give a simple example that demonstrates that the FT algorithm,
might not reproduce the parent distribution of a feature for a linear process.
We used a simple Gaussian white noise process with mean zero and
unit variance and evaluated the simple feature:

\beq \label{ableit}
   f({\bf x}) = \frac{1}{N-1} \sum_{t=1}^{N-1} \, |x(t+1) - x(t)|
\eeq

\marginpar {Please locate fig.1 near here}

for new realizations of the  process and for ensembles generated
by the FT algorithm. These ensembles were based on different realizations
of the process. The results were
independent from the number of data points $N$ used.
We only show the results for $N=16384$ in fig. \ref{surro_ex}.
To exclude any variance dependent effect, all time series are normalized
to mean zero and variance 1.
The lowest line
displays the cumulative distribution of the feature from $100$ new
realizations of the process. The lines above show the results for
$100$ realizations based on the FT algorithm. The
FT algorithm fails to recover the parent distribution. The surrogate
distributions $p(f|\surr)$ are grouped too tightly around the value of
the test statistic $f(\x)$ (marked by $\Diamond$)
for that time series on which the surrogates were based.

The Kolmogorov-- Smirnov-- test for Gaussianity does not reject
the consistency with a normal distribution at the $99 \% $ level
of confidence for the shown distributions. Thus, they
 may be described by mean and standard deviation.
The standard deviation of the true distribution is underestimated
by a factor of two. The Kolmogorov-- Smirnov-- test for consistency of
two distributions applied to each surrogate distributions and the distribution
of new realizations of the process rejected the null hypothesis in every case
at the $99 \% $ level of confidence.

The choosen test statistic is indeed able to detect nonlinearities since
its expectation value for a Gaussian white noise process with unit variance
is $2/\sqrt{\pi} = 1.12...$ and for ''gaussified'' data of the logistic
map, $r=4$, which shows also a flat spectrum, it is $1.09...$ .

\section {Linear Processes} \label{review}
The misallocation of $\mu_{\surr}$ and the underestimation of the true
standard deviation by $\sigma_{\surr}$
that appear in fig.~\ref{surro_ex} can be explained in terms of the theory of
linear processes.
\subsection {Time Domain}

 In order to model the fluctuations in the periodicity of sunspot data
Yule \cite{yule} introduced linear stochastic systems:

\beq
   x(t) = a_1 x(t-1) + a_2 x(t-2) + \epsilon(t),
        \quad \epsilon(t)\sim{\cal N}(0,\sigma^2) \quad .
\eeq

These are now called autoregressive (AR) processes can be
generalized to autoregressive moving average (ARMA) processes by
including not only past values of the observations but also past values
of the noise.  The general ARMA[p,q] process reads:

\beq
 x(t) = \sum_{i=1}^{p} a_i x(t-i) + \sum_{j=1}^{q} b_j \epsilon(t-j) +
   \epsilon(t)
\eeq

State space models are a further generalization.  They
consist of a vector valued AR[1] process and a linear observation
function with observation noise:

\begin{eqnarray}
   \vec{x}(t) & = & {\bf A} \vec{x}(t-1) + \vec{\epsilon}(t)\nonumber\\
         && \vec{\epsilon}(t) \sim {\cal{N}} (0,{\bf Q}) \\
   y(t)   & = & {\bf C} \vec{x}(t) + \eta(t)\nonumber\\
          && \eta \sim {\cal{N}}(0,{\bf R})\nonumber \quad,
\end{eqnarray}

where $\bf{Q}$ is the covariance matrix of the dynamical noise, and $\bf{R}$
the variance of the observation noise.  State space models model
observation noise explicitly.  For AR and ARMA models, observation
noise causes the estimated parameters to be biased towards zero.

Once the type of model has been chosen, one must select the model
order. Generally, one fits a
sequence of higher and higher order models and tests the prediction
errors for consistency with white noise, e.g. by a Kolmogorov --
Smirnov -- test on the periodogram \cite {brockwell}.

If a particular class and order of model is selected, when the
parameters of the model are estimated using data from a random
process, the estimated parameters are random variables. By
restricting the number of parameters to be fit, one insures that as
the number of observed data increases, the variance of the estimates
goes to zero. In the next subsection we emphasize that periodograms
do not have this property.

\subsection {Frequency Domain}

We now examine the spectrum of linear processes and the relationship between
the spectrum and the periodogram.
The autocovariance function $\ACF(t)$ is given by

\[
\ACF(\tau)  \equiv \left<x(t) x(t+\tau) \right> \quad ,
\]

and the spectrum is defined by:

\begin{eqnarray}
S(\omega) &\equiv&  \sum_{t=-\infty}^{\infty} \ACF(t) e^{-i\omega t}\nonumber
  = \left<|\frac{1}{\sqrt{N}}\sum _t x(t) e^{-i\omega t}|^2\right>\nonumber\\
    &=& \left<|x(\omega)|^2\right>\nonumber
   =  \left<\Per(\omega)\right>\nonumber \qquad .
\end{eqnarray}
%
The spectrum of a state space model is given by :

\beq \label{specar}
      S(\omega) = {\bf C} ({\bf 1}-{\bf A}e^{-i \omega})^{-1} {\bf Q}
         ({\bf 1}-{\bf A}e^{i \omega})^{-1\,T} {\bf C}^T + {\bf R} \quad.
\eeq
Thus the spectrum $S(\omega)$ of this process is a smooth function of
$\omega$. Spectra of AR or ARMA processes are special cases
of eq. (\ref {specar}).

We now turn to the distribution of the periodogram.  For details see
\cite{brockwell,priestley}, we only summarize the
results. Asymptotically, the Fourier transform of a linear stochastic process
\beq
x(\omega) = \frac{1}{\sqrt{N}} \sum_t  x(t) \cos(\omega t) +
            \frac{i}{\sqrt{N}} \sum_t  x(t) \sin(\omega t)
\eeq
is a complex Gaussian random variable
\beq   \label{fourverteil}
x(\omega) =  {\cal {N}}(0,\frac{1}{2}S(\omega)) +
               i {\cal {N}}(0,\frac{1}{2}S(\omega))
\eeq
where $S(\omega) $ is independent of $N$. Furthermore, Fourier
components are asymptotically uncorrelated for different frequencies.
In terms of modulus and phase instead of real and imaginary part,
this means that not only is the phase a random variable but so is the modulus.

The distribution of the periodogram
\beq
\Per(\omega) = \left(\frac{1}{\sqrt{N}}\sum_t x(t)\cos(\omega t) \right)^2
              + \left(\frac{1}{\sqrt{N}} \sum_t x(t) \sin(\omega t)\right)^2
\eeq
 follows from eq. (\ref{fourverteil}) to be a $ \chi^2 $ distritubion with
 two degrees of freedom $ \chi^2_2 $:
\beq
\Per(\omega) \sim  \frac{1}{2}S(\omega)\, \chi^2_2
\eeq
again independent of $N$.

Since the mean and variance of $\chi^2_2$ are $2$ and $4$
respectively, the standard deviation of the periodogram is equal to its
 mean, \ie
\[
   \Per(\omega) = S(\omega) \pm S(\omega)  .
\]
Thus, the periodogram fluctuates wildly regardless of the number of
data points used to calculate it.  It is an unbiased but not a
consistent estimator of the spectrum since its variance does not
decrease with $N$. This is true not only for linear stochastic processes
but also for nonlinear stochastic and even for chaotic processes if only
the autocovariance function decays fast enough.

\section {Consequences for the Surrogate Data Method} \label {consequ}

The characteristics of the spectrum and
periodogram described in the previous section lead to the following
two points concerning the FT algorithm:
\begin{enumerate}
\item The FT algorithm does not use its estimate of the spectrum
correctly.
Both phases  and amplitudes should be randomized.
\item A better estimator of the spectrum than the periodogram is required.
\end{enumerate}
Below we elaborate on these points.

\subsection{Randomizing Phases and Amplitudes} \label{randi_all}
       Eq. (\ref {fourverteil}) means that at each frequency the
       Fourier transform of a time series from a linear stochastic
       system is a two
       dimensional Gaussian random variable. The FT algorithm, by
       fixing the modulus and only randomizing the phases, restricts
       the new realization to a set of measure zero in the set of all
       possible realizations of the linear stochastic system under
       consideration.
       It interprets the periodogram of a stochastic process
       as the spectrum of $N/2$ linear
       deterministic oscillators with fixed amplitudes.
       Thus, instead of testing for linear stochastic systems, the FT algorithm
       by construction tests for the time -- discrete version of the
       following deterministic process $y(t)$:

  \begin{eqnarray}
   \ddot {x}_i(t) &=& - \omega_i^2 x_i(t), \quad x^2_i(0) +
      \left(\frac{\dot{x}_i(0)} {\omega_i}\right)^2 = Per(\omega_i) \nonumber\\
    y(t)&=&\sum_{i=1}^{N/2} x_i(t)  \qquad .
  \end{eqnarray}

Note, that this process depends on the length $N$ of the time series.

        If the true spectrum of the process were known the following
         algorithm would yield the correct distribution of surrogates.
         The  algorithm is based on eq. (\ref{fourverteil}) that
         connects the spectrum $S(\omega)$ with the variance of the
         complex random variable $ x(\omega) $.
\begin{itemize}

  \item For each frequency $ \omega_i=\frac{i}{N\Delta t}, \, i=1,
          ..., \frac{N}{2} $, draw two Gaussian distributed random
          numbers, multiply them by $ \sqrt{\frac{1}{2}S(\omega_i)} $
          and take the result as the real and imaginary parts of the
          Fourier transform of the desired data.

  \item For reasons of symmetry the last frequency is always real for
        even number of data points. Thus, in this case, only one
        Gaussian distributed random number has to be drawn.

\item To obtain a real valued time series, chose the Fourier components
        for the negative frequencies as the complex conjugates of the
        components at positive frequencies, i.e.,
        $x(-\omega)=x^{\ast}(\omega)$.

  \item Do an inverse Fourier transformation of $ x(\omega) $ to get
        the surrogate time series.

\end{itemize}

\marginpar {Please locate fig.2 near here}

Fig. \ref{surro_wahr} shows the results of a simulation study analogously
to fig. \ref{surro_ex}. Again, the bottom line displays the cumulative
distribution of the feature from eq. (\ref{ableit}) for new realizations
of the white noise process. The upper lines show the distributions based
on the proposed algorithm using the true flat spectrum of white noise.

\subsection{Estimating the Spectrum} \label{est_spec}

The critical issue for the procedure described in the previous section
is spectral estimation.  There are two general approaches commonly used
for this purpose: Fit an AR or ARMA model to the data and calculate
the spectrum from the resulting parameters  or smooth the periodogram.
For more than the brief discussion given here, see
\cite{brockwell,priestley}.

\subsubsection*{ Spectral estimation by fitting AR or ARMA models}

  This method cannot be recommended for two reasons. On the one hand,
  as discussed in section \ref{review},  in presence of
  observation noise these models do not give correct results even
  if the underlying dynamic is linear.  On the other hand, if the process
  is nonlinear they are, in  general, not consistent estimators
   of the spectrum. Of
  course, the parameters of linear models determine the spectrum. But
  not every spectrum, e.g. of a chaotic process, is well described by a
  linear model, since the prediction errors will not be consistent
  with white noise.  Increasing the order of the model does not help,
  because the ``spectrum'' calculated from the parameters will begin
  to interpolate the periodogram. For the same reason, surrogate data
  generated by realizations from a fitted ARMA model test, in general,
  for some linear model, but not for that specified by the spectrum.

\subsubsection*{ Estimation by smoothing the periodogram}
  Estimating the spectrum by smoothing the periodogram rests on three
  results of the theory of linear processes: the expectation value of
  the periodogram is the spectrum, the spectrum is a continuous
  function, and the random variables $\Per(\omega)$ and
  $\Per(\omega')$ are asymptotically uncorrelated. The idea
  is to smooth the periodogram to reduce the variance (or equivalently
  split the time series into segments and average the periodograms).
  Denoting the number of data points by $ N $, the number of frequency
  bins included in the smoothing by $n_s$, one can show that in the
  limit of $ N \rightarrow \infty$ and $ n_s \rightarrow \infty$, with
  $n_s$ constrained to increase slowly enough to insure $\frac{n_s}{N}
  \rightarrow 0$, this procedure leads to a consistent estimator for
  the spectrum.

For finite data sequences, the smoothing will introduce a bias and
there is a trade--off between bias and variance of the estimator which
is discussed at length in \cite{brockwell,priestley}.

\subsubsection* {Surrogates based on an estimated spectrum}

The bias and the variance of the spectral estimate lead to
a misspecification of the surrogate data distribution that
vanishes asymptotically.

\marginpar {Please locate fig.3 near here}

Fig. \ref{right_ex} is analogous to fig. \ref{surro_ex}.
It shows the results of a simulation study
using a moving average kernel of width 2000 frequency bins to
estimate the spectrum and the algorithm of section \ref{randi_all}
to generate the surrogate data. The Kolmogorov - Smirnov - test
for consistency of two distributions does not reject the null
hypothesis in 7 of the 9 cases shown at the $99 \%$ level
of confidence, demonstrating the convergence.

\section {Conclusion}

The appeal of the general idea of surrogate data
testing is clarified by a comparison with the tests for
linearity proposed in the stochastic systems literature. These tests
rely on asymptotic results for linear processes, mainly
concerning third order properties. There are two classes of these
tests. The first class is based on fitting an AR model to the data
and investigating third order properties of the prediction errors.
Examples of these parametric tests are given by
\cite{keenan,tsay}.  The other class of tests is based on the
asympotitic property that the three--point--correlations:
\beq
  c(\tau_1, \tau_2) = \sum_t x(t) \, x(t+\tau_1) \, x(t+\tau_2)
\eeq
are zero for any linear process.
For examples of these nonparametric
tests see \cite{subba,hinich}. In order to perform the test,
$c(\tau_1, \tau_2)$ is Fourier transformed and the normalized
bispectrum is estimated. For linear Gaussian processes it should be
zero and the consistency with zero can be tested. Both classes of
tests test data of a given process by a property that is fulfilled by
all possible linear processes.

The intuition behind the surrogate
data method is that only that linear process that, jugded by its spectral
properties, could have generated the data needs to be tested.

The sampling properties of a test statistic of a linear process depend
on the the details of the process and the statistic. In particular,
the variance will often be larger for processes with sharper peaks in
the spectrum or equivalently longer coherence times. For example, the
variance of the sample mean $\hat {\bar {x}}$ for an iid Gaussian process
${\cal N}(0,\sigma^2)$ is $\frac{\sigma^2}{N}$, but if the process has
correlations, the variance of the sample mean will decrease more
slowly as a function of $N$. By construction, the surrogate data
method produces the process specific sampling properties.

The power of any test, i.e. the ability to detect a violation of
the null hypothesis, depends on the statistic. The bispectral tests
considering the three--point--correlations are sensitive to quadratic
nonlinearities but can be fooled completely by processes showing only
cubic nonlinearities like the Duffing - oscillator. The general
scheme of surrogate data testing can use any statistic, and the power
of the test will depend on the statistic chosen. Thus one can detect
a larger class of violations of the null hypothesis.

We showed by an example that the FT algorithm proposed by Theiler et al.
\cite{theiler1} might specify the distribution of
the test statistic incorrectly, independent of the number of data points used.
Based on the theory of linear stochastic systems,
we proposed an alternative that involves spectral estimation.
Therefore, testing for linearity by the idea of surrogate data is, in general,
also a procedure that is only valid asymptotically.
For a finite number of data, simulation studies should
be performed to evaluate the estimation error.

\vskip 2cm
Acknowledgements: \\
I like to thank A. Fraser for many clarifying discussions and
J. Theiler for comments on an earlier version of this paper.

\clearpage

{\bf Figure captions}

\begin{figure*}[h]
\caption{Cumulative distributions based on 100 realizations for the test
    statistic $f(\x)= \frac{1}{N-1}
   \sum |x(t+1)-x(t)|$. The line at the bottom displays the distribution for
   new realizations of the process. The lines above show the distributions for
  the FT algorithm  based on different realizations of the process.
 The values of the test statistic $f(\x)$
  for the realizations which were surrogated are marked.
The most extreme values of the cumulative distributions are not plotted
for clearness.
}
\label{surro_ex}
\end{figure*}

\begin{figure*}[h]
\caption{Cumulative distributions based on 100 realizations for the test
    statistic {$f(\x)= \frac{1}{N-1}
   \sum |x(t+1)-x(t)|$}. The line at the bottom displays the distribution for
   new realizations of the process. The lines above show the distributions
   based on the algorithm of section 4.1 using the true spectrum.
The most extreme values of the cumulative distributions are not plotted
for clearness.
}
\label{surro_wahr}
\end{figure*}

\begin{figure*}[h]
\caption{Cumulative distributions based on 100 realizations for the test
    statistic {$f(\x)= \frac{1}{N-1}
   \sum |x(t+1)-x(t)|$}. The line at the bottom displays the distribution for
   new realizations of the process. The lines above show the distributions
   based on the algorithm of section 4.1.
  The values of the test statistic $f(\x)$
  for the realizations which were surrogated are marked.
The most extreme values of the cumulative distributions are not plotted
for clearness.
}
\label{right_ex}
\end{figure*}

\clearpage

\begin{figure}
\centering
\epsfxsize=14cm
\epsfbox{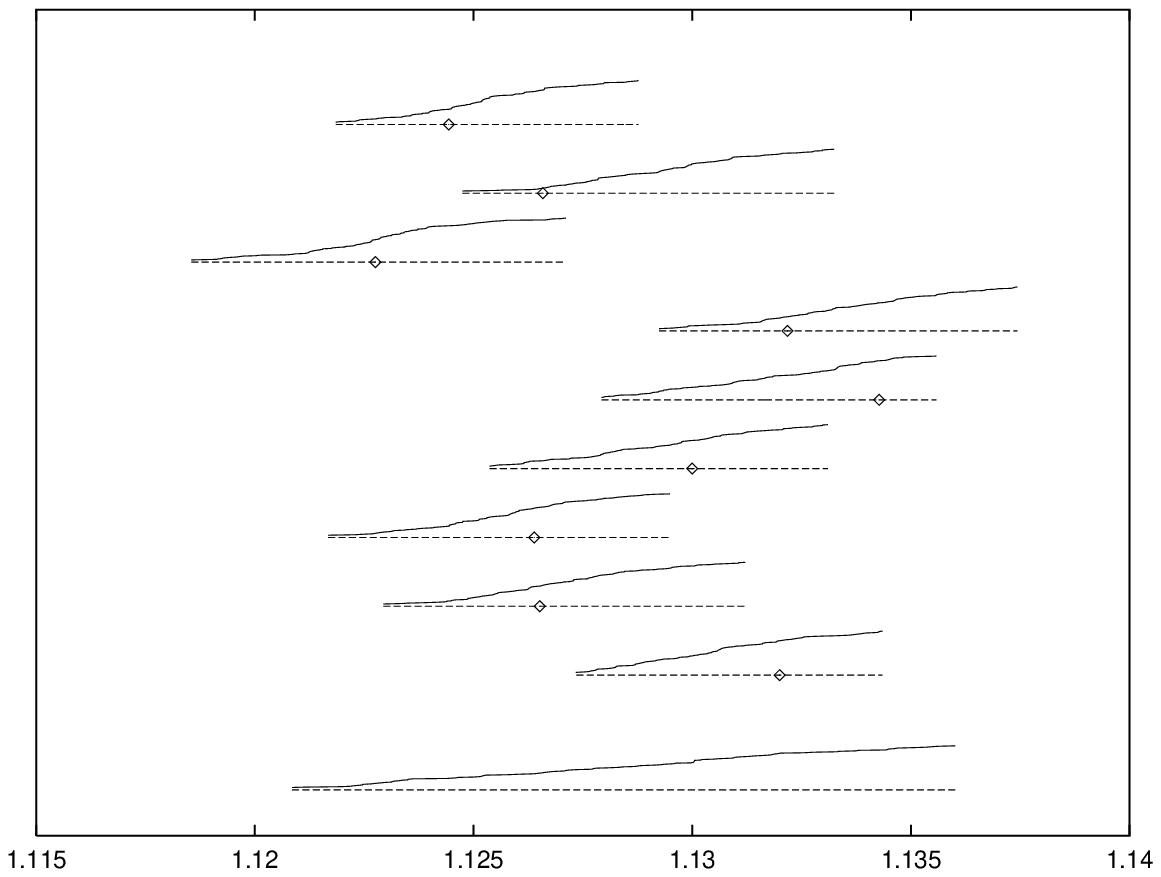}
\end {figure}

\vskip 1cm
\begin{center}
Fig. 1
\end{center}

\clearpage

\begin{figure}
\centering
\epsfxsize=14cm
\epsfbox{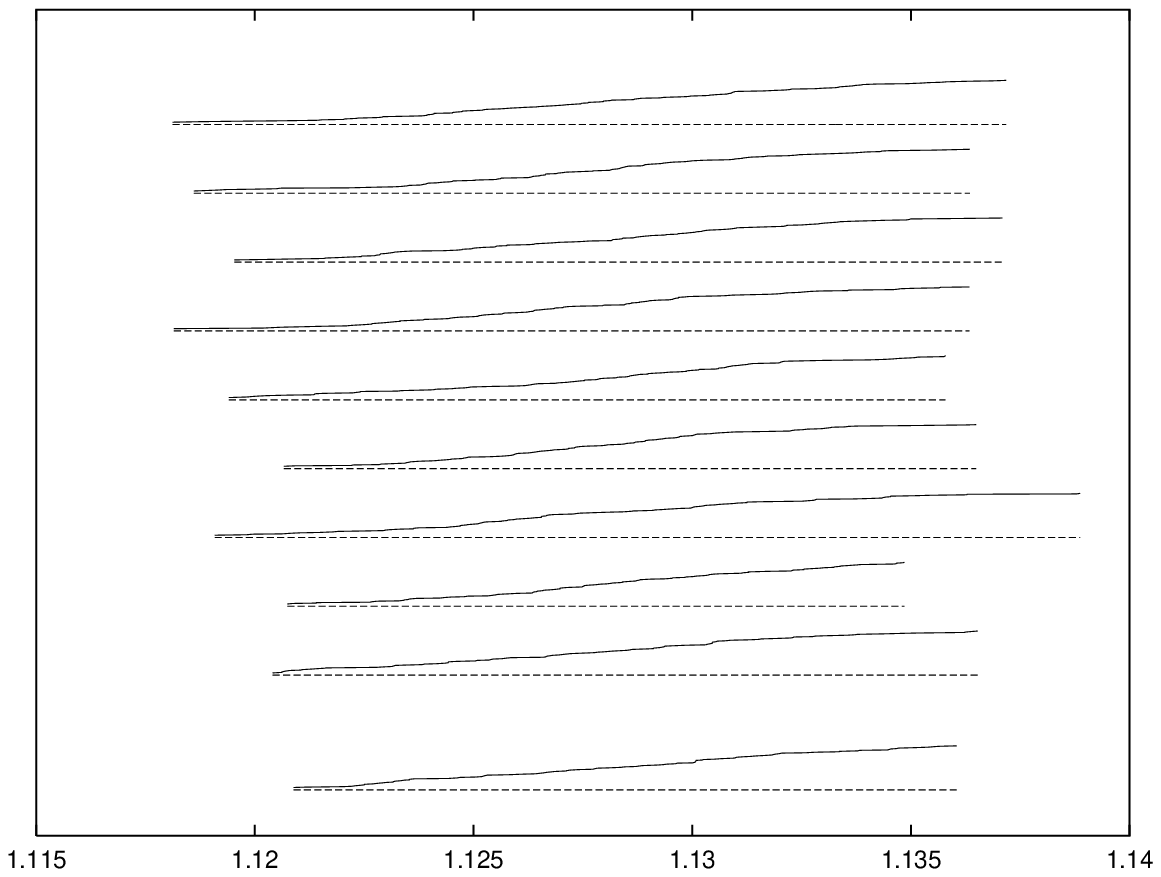}
\end {figure}

 \vskip 1cm
\begin{center}
Fig. 2
\end{center}

\clearpage

\begin{figure}
\centering
\epsfxsize=14cm
\epsfbox{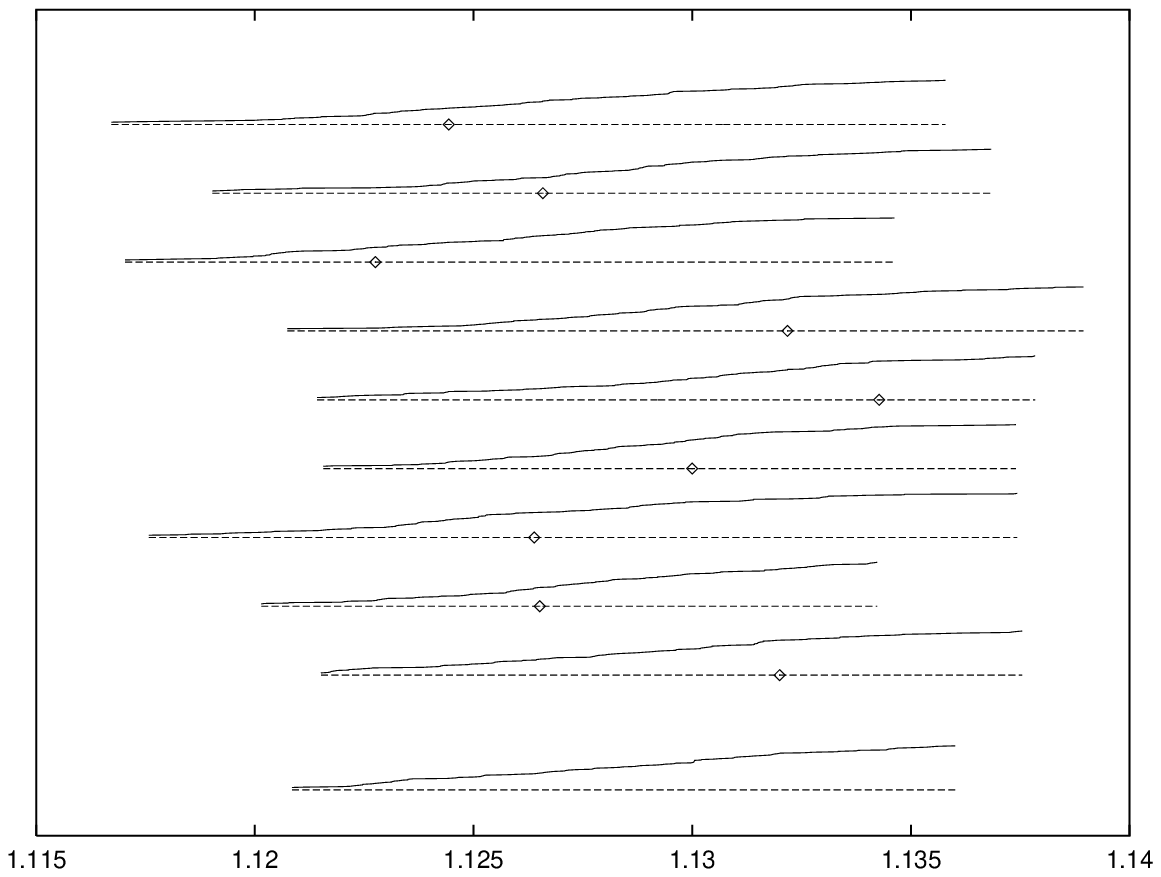}
\end {figure}

 \vskip 1cm
\begin{center}
Fig. 3
\end{center}

\end {document}